A.D. Avesov, M.A. Belov, V. Chaloupka, W.M. Dougherty, A.A. Gafarov, D.Hutcheon, A. Khugaev, B.Khurelbaatar, Yu.N. Koblik, R.Kh.Kutuev, G. Liang, H. Lubatti, V.V. Mialkovski, D.A.Mirkarimov, G.A. Mkrtchyan, G. Musulmanbekov, V.A.Nikitin, P.V.Nomokonov, A.V. Pavlyuk, G.A. Radyuk, I.A. Rufanov, W.G. Weitkamp, S. Yen, B.S. Yuldashev, T.C.Zhao, R.Ya. Zulkarneev.

# MASS SPECTRUM MEASUREMENT IN THE REACTION (p, 2p) AT PROTONS ENERGY 500 MeV ON NUCLEI C, Al, Cu AND Pb

## 1. Introduction.

The research of abnormal phenomena in the spectra of the particles effective mass in nucleon-nucleon and nucleon-nucleus interactions is of great interest. To the present time in scientific centers of CIS, France, Germany, USA, Japan and etc. a lot of studies with the purpose of baryon structure discovering in the two-nucleon system have been executed. Obtained results, however, are not convincing because of the presence of both numerous indications on dinucleon resonances existence in the wide range of effective mass (from the sum of two nucleon masses to the value ~ 2.9 GeV), and experiments, which do not confirm the production of such structures in nucleon systems [4]. These facts force us to search for a new possibility for discovering of dibaryon resonance [5] and to make higher demands to the measurement conditions in traditional experiments.

With the purpose of dibaryon structures research in mass spectra of the reaction A(p, 2p)X, whose investigation were carried out at the energy of polarized proton beam 500 MeV in the experiment №627 ("A search for narrow resonant-like structures in pp and pA interactions using the polarized proton beam at TRIUMF"[6]), correlation spectra of secondary protons at the simultaneous irradiation of nuclei targets of carbon, aluminum, copper and lead were measured. In the present work some of obtained results are discussed.

## 2. Experimental plant and measurement conditions.

Experimental plant [7–8] is the nonmagnetic scintillation spectrometer (Fig.1), located on removed beam of the TRIUMF cyclotron of the National meson laboratory of Canada. The spectrometer has two movable arms with eight ΔE·E telescopes for the particle type identification and their energy determination. The range of covered angles is 32°–148°. The angle spread between centers of nabouring telescopes is 14°, and the spatial angle of each telescope is 0.012 sr.

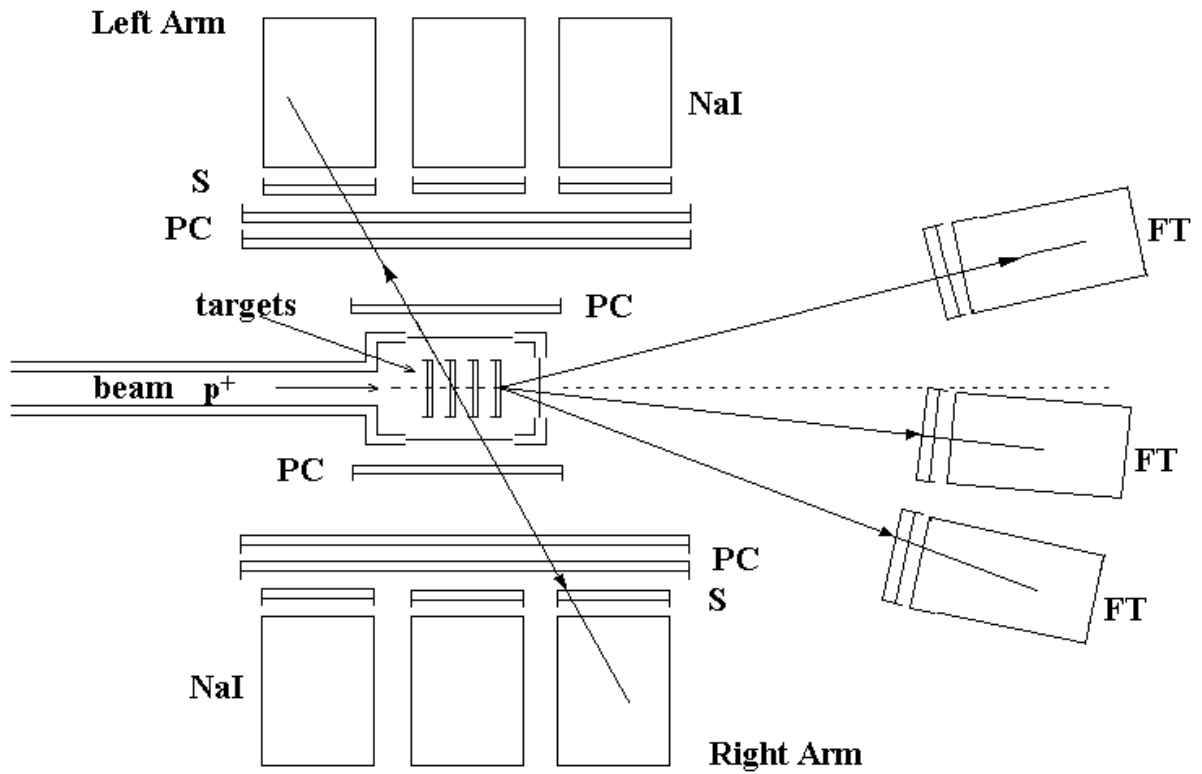

Fig.1. Schematic view of scintillation spectrometr set-up at polarized proton beam of TRIUMF cyclotron.

PC – multiwire proportional chambers.
NaI – sodium iody scintillation counters.
S - disc scintillation counters.
FT – forward telescopes.

Plastic scintillators with diameter 100 mm and thickness 10 mm are used as ΔE-detectors in all telescopes, and NaJ(Tl) crystals with dimensions 150 mm × 150 mm play a role of E-detectors. There are three multithread proportional cameras (MPC) on each movable arm: X and Y cameras with step 2 mm and dimensions 512 mm × 384 mm in front of ΔE counters at the distance 495 mm and 520 mm from targets respectively, and one X camera with step 3 mm and dimensions 192 mm × 70 mm directly on vacuum target box at the distance 90 mm from the beam. Besides telescopes, located on movable arms, extra three stationary plastic (ΔE·E)-telescopes, placed on angles and "towards": Θ=10.9°, 21.7° and 30° are used in the structure of spectrometer. Spatial angles of these telescopes are 0.001, 0.002 and 0.003 sr. The angle allowance of the plant is 0.5°.

As the using of a number of MPC on each arm permits us to define particle tracks reliably, the simultaneous work under the beam on several targets with different composition becomes possible. For this purpose the target block was placed into the vacuum chamber, which is the hermetic capacity, where lateral and back walls have thin lavsan windows. Two target groups (CH, Al, Cu, Pb for the main measurements and $CH_2$, $CD_2$ for calibration) were fastened to a frame, which moved remotely "up and down" inside the target block and brought necessary target group into the beam. The target thickness (in mg/sm$^2$): C–41.5, Al–64.9, Cu–77.7, Pb–64.8. Simultaneous irradiation of the section of several targets permitted us to reduce the experimental time and measurement mistakes. Electronics and TRIUMF standard system of data accumulation during on-line mode measuring on VAX computer provided reading of the information from MPC, scintillation counters and receiving of the target signal and signal digitizing from every counter. In each event of the main trigger the signal amplitude from a counter, the time between trigger operation and a signal from a shaper of every counter and information from MPC without any selection and compression in fixed format was recorded from all counters. The operation time of each ΔE counter relatively to trigger served as for reducing of the casual concurrence background, caused by the bunch structure of a beam, so and for identification of soft particles, stopped at 1 sm from a plastic of ΔE.

For the control of NaJ(Tl) counters work and respective analog electronics light-diodes were used, on which momenta of five different amplitudes were given consistently under the computer control and their signal amplitudes were recorded into the data set of separate trigger. It permitted us to use for data treatment relative changes of the difference of E-counter signal amplitudes from light-diodes as the scale change of an analog-digital converter for events from the main trigger.

A proton energy in the measurements was 500 MeV, beam current– (0.15–0.2) nA. Beam polarization (~20–40% in one seance) was changed on "up", "down" and "zero". Polarization measurements were carried out every three minutes with the help of standard polarimeter, recorded recoil protons from pp-scattering in $CH_2$-target, placed in front of the spectrometer. The observation for beam location during measurements was realized via TV-monitors, registered a luminescence 300 mkm of ZnS-pellicle, located before the target block.

The estimation of possibility of event superposition was conducted at beam current 0.1 nA. The interaction number in one trigger (0.07 g/sm$^2$) at allowance time of coincidence in trigger scheme 100 ns (2 bunches) is $10^{-2}$ and the possibility of superposition is too little. A loading of E-counters does not exceed $2·10^4$ events per second, that is why the possibility of superposition in each counter during the strobe time is also too little. Plant allowance on mass ~ 4 MeV at candlepower ~ 0.2 sr [8].

## 3.   Measurement results.

In the present work two-proton spectra, received from particle coincidences, got into telescopes, placed on opposite moveable spectrometer arms, were measured.

Belonging of the secondary part track to one or another target was determined by MPC coordinates and stable location of primary beam relatively to researched targets. At treatment only that tracks were taken into account, which passed through one of the targets and ΔE-counters of the telescope. If more than one track passed through ΔE-counter, the event was rejected as defective.

Two-proton mass spectra, received at conditions of summarizing of all correlated pairs of pp-interactions, accumulated during the experiment, are shown on Fig.2. Here (and further) the range of overlapping effective masses stretches from 1954

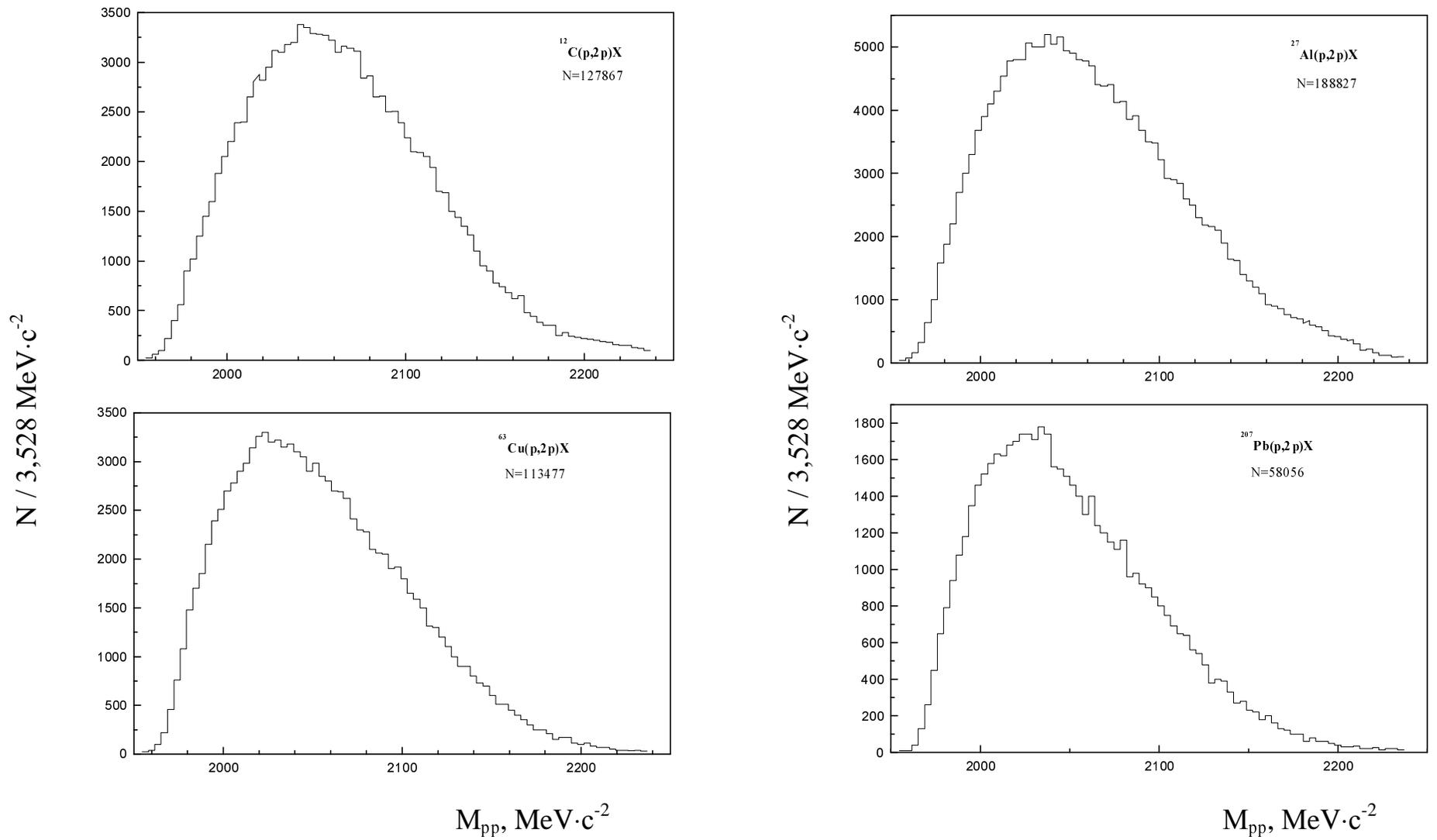

Fig.2. Total mass spectra of proton pairs for C, Al, Cu and Pb nuclei.

MeV/c$^2$ to 2220 MeV/c$^2$ at mass step 3.528 MeV/c$^2$. Summary statistic in effective mass spectra is large and for carbon– 127867, for aluminum– 188827, for copper– 133477 and for lead– 58056 events.

As it follows from the figure the spectra nature is smooth and any anomalies is not observed. It can mean, those effects, joined with the resonance structures production, are absent or they are commensurable with another (background) processes (for example, quasi-elastic pp-scattering, scattering on nucleon clusters, cascade processes, etc.) at such a degree, that at given conditions of experiment and treatment they are not displayed. However, the situation becomes different if we realize a selection of complanar proton pairs from received spectra and limit the registration angle in the reaction plain only by one telescope on each spectrometer arm (14º).

Mass spectra of proton pairs for such a selection, carried out separately for each researched target, are shown in Fig.3. The total event number in spectra is 10886, 13024, 5518, 5313 for carbon, aluminum, copper and lead respectively. At these conditions mass spectra have noticeable irregularities. The most significant of them are shown on figure by arrows with numeral mass values. Their comparison with the available data form literature (see, for instance, final figure-scheme 31 in report [3]) indicates the fact, that observable anomalies in measured spectra are comparable to experimental data on dinucleon resonances. Mass values, concerned to them, display on researching nuclei with one or another trustworthiness degree, and to wit– 1998, 2008, 2017, 2028, 2046, 2067, 2087, 2096, 2106, 2156 MeV/c$^2$. An anomaly with the mass 1998 is observable only in spectra of copper and lead and another one with the mass 2096 is the most noticeable on lead. Anomalies with the mass 2008 and 2028 are not displayed on copper, and the one with the mass 2017 is not revealed on carbon. States with masses 2046, 2067, 2087 and 2106 are manifested in all spectra, and the one with the mass 2156– only in copper spectrum.

Mass spectrum (histogram 6), summed up on all investigated nuclei, are shown in Fig.4. Histograms 1–4 are plotted on data from Fig.3 for lead, copper, carbon and aluminum respectively. Statistical provision in total spectra is 34740 events (statistical criterion on [3]– 815). In this figure curve 5 is the background, approximated by the

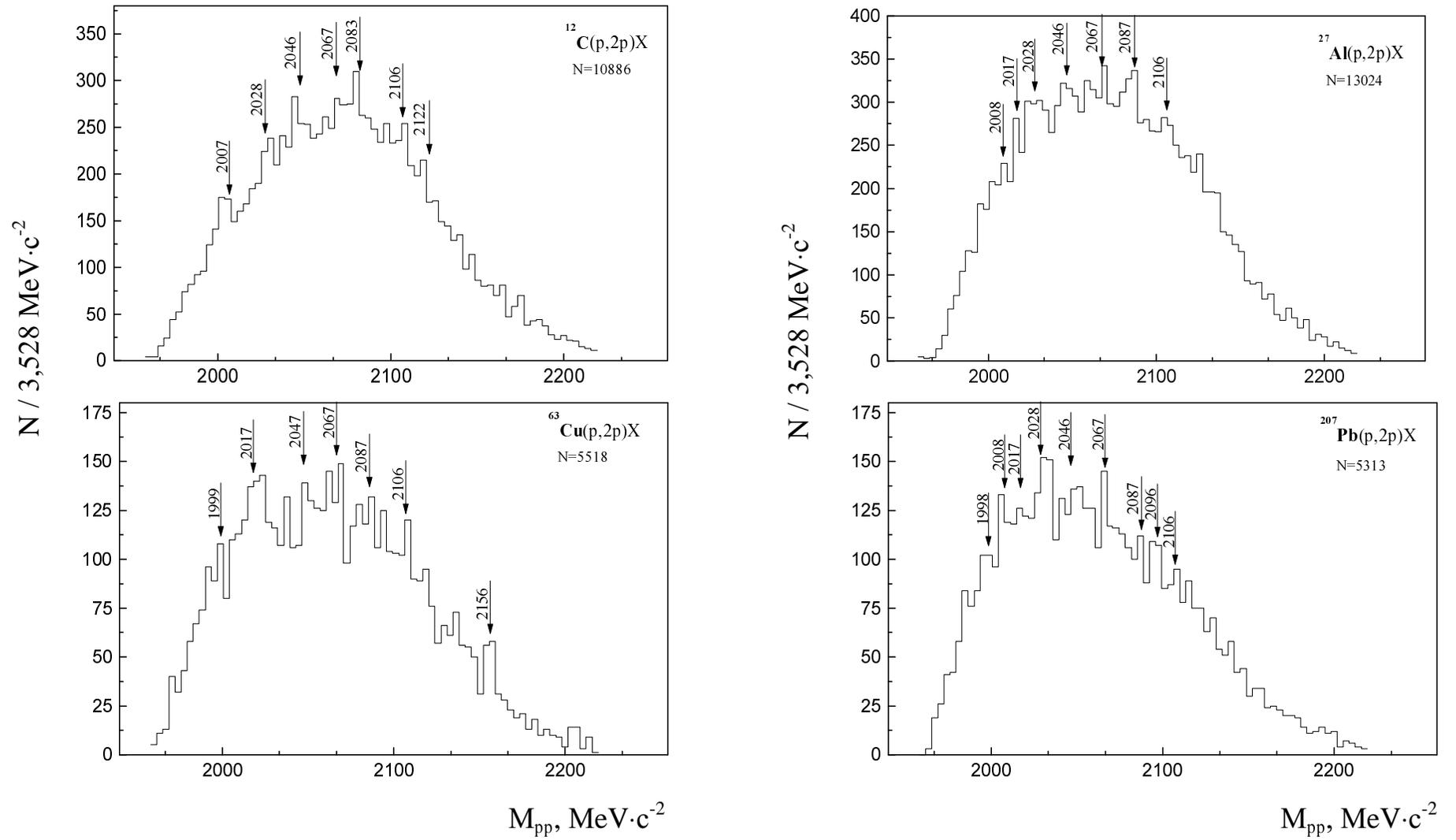

Fig.3. Mass spectra of complanar proton pairs for C, Al, Cu and Pb nuclei.

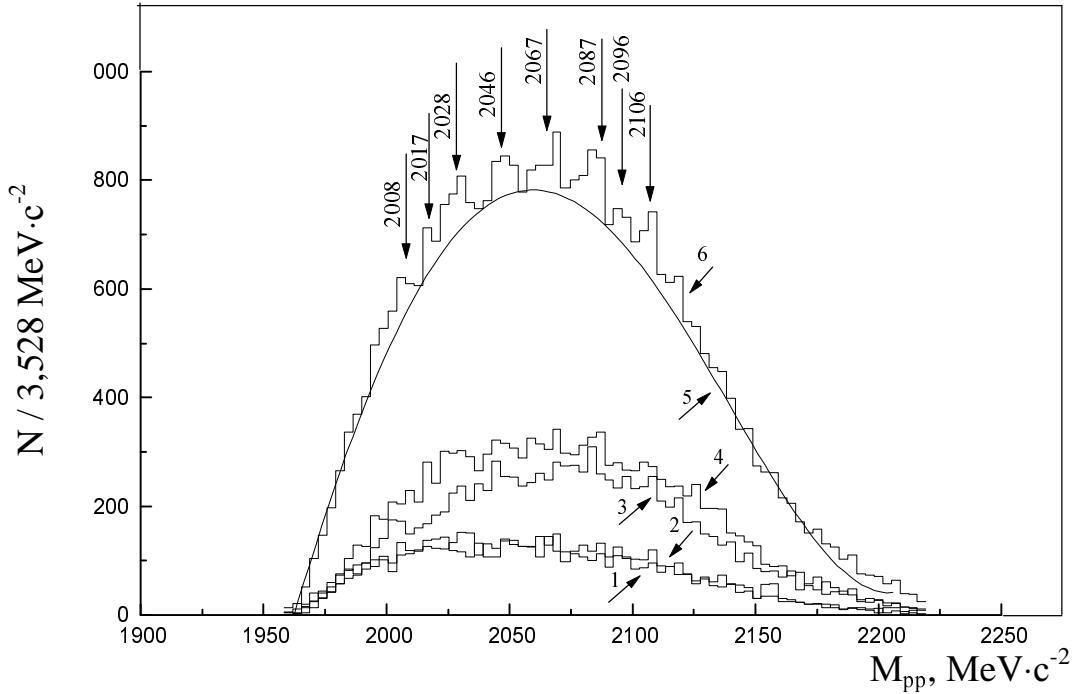

Fig.4. Total mass spectrum of complanar proton pairs for C, Al, Cu and Pb nuclei.

following 4-th degree polynomial:

$$N_f = 2{,}03 - 4{,}04 \cdot 10^4 \cdot M_{pp} + 30{,}02 \cdot M_{pp}^2 - 0{,}01 \cdot M_{pp}^3 + 1{,}22 \cdot 10^{-6} \cdot M_{pp}^4$$

Available anomalies overlap the level of background distribution on 2.5–3 statistical mistakes relatively that curve. Mass values of displayed anomalies are 2008, 2017, 2028, 2046, 2067, 2087, 2096 and 2106 MeV/$c^2$. A mistake in mass value determination of anomaly is ±4 MeV.

## 4. Results discussion.

As the main information about anomalies, mentioned previously, consists only of mass values, it is difficult to realize any classification. In the work [9] the possible classification of dibaryon resonances and their production mechanism was offered. In this classification all resonances are divided on three classes: NN, NNπ and ΔN-dibaryons. Each dibaryon group consists only of revolving belts, based on the sum of particles masses from corresponding class. Fig.5 represents this mass diagram. Possible candidates in dibaryon resonances, discovered by us, in accordance with the scheme under discussion, are divided on two groups. The first group of resonances with masses 2008, 2028, 2067 and 2087 MeV/c$^2$ lays well on rotation belts of NN-resonances. The second group of three resonances with masses 2017, 2046 and 2106 MeV/c$^2$ is the rotation belt of NNπ-resonances.

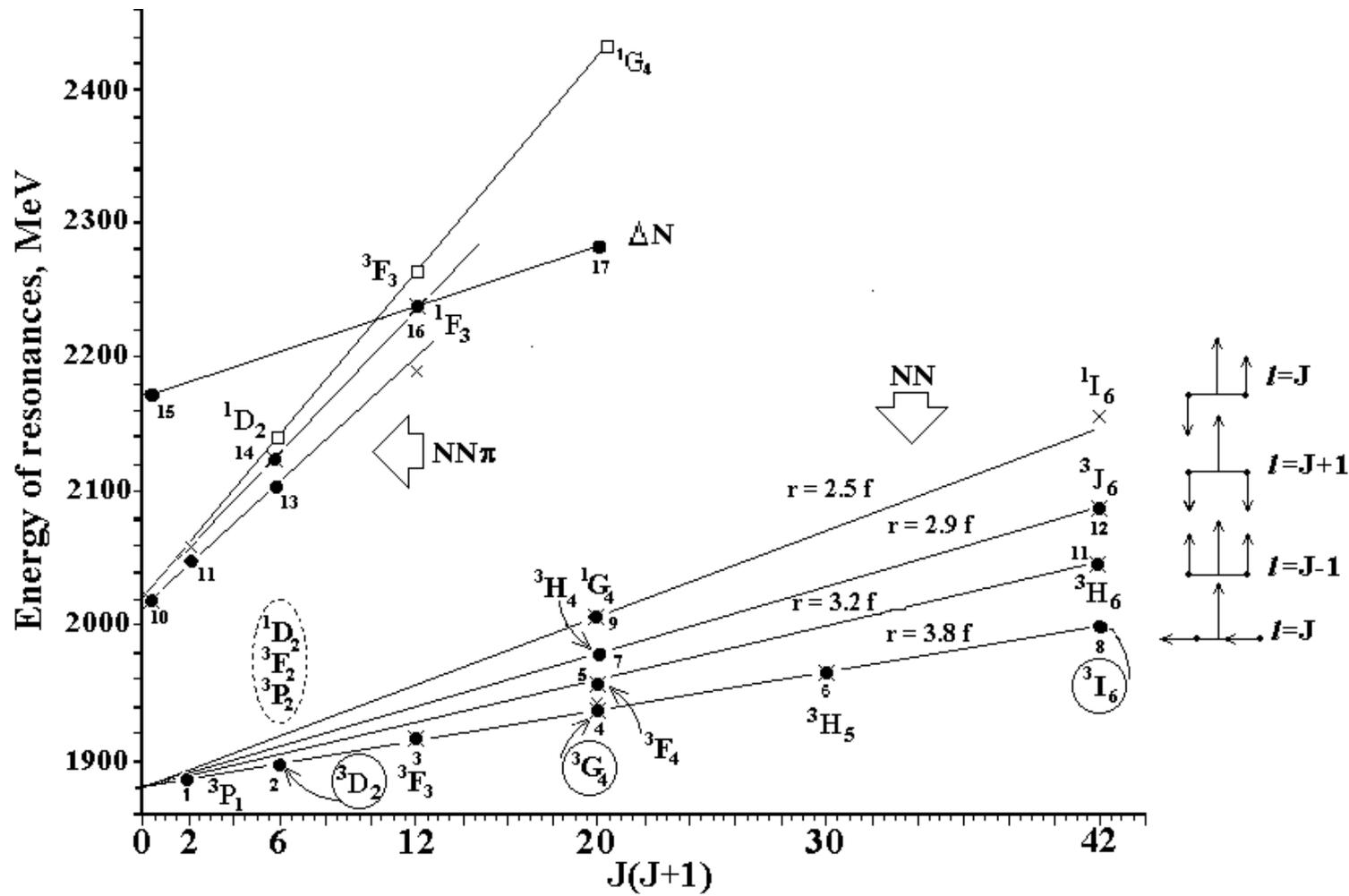

Fig. 5 Revolving belt groups of dibaryon resonance NN, NNπ and ΔN. Resonances in order of mass increasing are marked by figures. Mutual orientations of orbital (long arrows) and spin (short arrows) moments are shown for NN-belts.

Despite the fact, that not all of obtained mass values agree with the systematization [9], it seems quite reasonable for us. In supposition [9] resonances are formed at rather big internucleon distance of an order 2.5–4 f on the slope of interaction potential in local irregularities, which can appear as a result of nucleon spins overturn. The fact, that resonances form rotation belts, pints on making of irregularities at definite distance between nucleons but the system inertia moment remain constant. The explanation of the spin orientation change at fixed distance, where nuclear interaction is weak, electromagnetic forces set out spins in one of possible positions. The energy of electromagnetic spin interaction with magnetic field, formed by moving charges, increases as R$^{-3}$. As nucleons drawing together, nuclear forces become apparent. A part of nuclear potential, responsible for orientation change, increases faster than R$^{-3}$ and at definite distance becomes equal to the energy of electromagnetic interaction. At this distance nuclear spin-orbital interaction changes spin orientation.

The energy of nucleon electromagnetic interaction, responsible for spin orientation is:

$$U_{elm} = (\vec{\mu}_S \vec{H}_L) = g \frac{e^2}{mc^2} \cdot \frac{\hbar^2}{mR^3} (\vec{l}\vec{s}) \ .$$

Phenomenological nucleon-nucleon potential, including spin-orbital component, has been taken from [10] (Hamada-Jonson potential):

$$V_{ls} = 2{,}77 \cdot 3{,}65 \cdot \left[\frac{exp(-x)}{x}\right]^2 (\vec{l}\vec{s}) \ ,$$

where $x = \dfrac{R}{1{,}43\,\text{fm}}$, figures are expressed in MeV and are taken for triplet even states.

A behavior of $U_{elm}$ and $V_{ls}$ is shown in Fig.6. For absolute values they become equal each other at internucleon distance R=4.54 f; this fact to a good degree corresponds to values, determined during the forming of dibaryon resonance systematization [9].

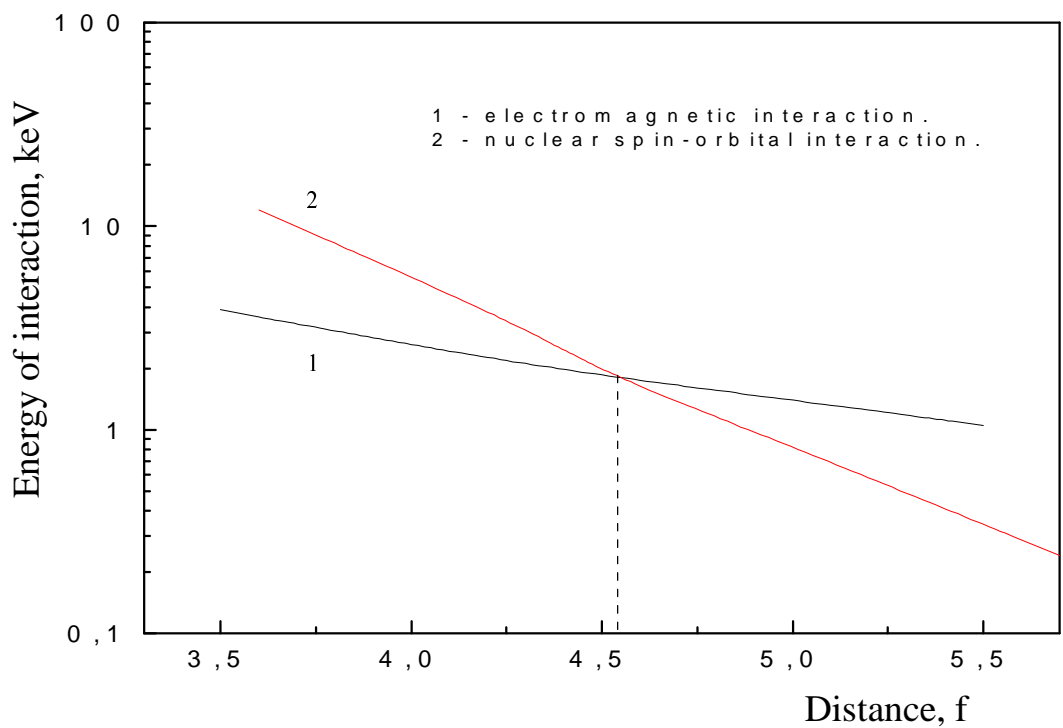

Fig.6. Comparison of the energy of electromagnetic interaction to the nuclear (spin-orbital) interaction.

## 5. Conclusion.

As our measurements of nucleon-nucleus interaction spectra show, displaying anomalies in most cases can be compared to observable dibaryon resonances. However, anomalies in mass spectra are presented only at the observance of some conditions (complanarity, narrow angle spread, etc.). for the receipt of information with the most trustworthiness, extra measurements with taking into account of these and others experimental features are needed.